\documentstyle[12pt,aaspp4,tighten]{article}
\begin{document}
 
\title{EXTREMELY METAL--POOR STARS. VII. THE MOST METAL-POOR DWARF,
CS~22876--032}

\vspace{1cm}

\author{John E. Norris}
 
\affil{Research School of Astronomy \& Astrophysics, The Australian National
University,\\ Private Bag, Weston Creek Post Office, ACT 2611, Australia;
jen@mso.anu.edu.au}

\author{Timothy C. Beers}

\affil{Department of Physics \& Astronomy, Michigan State University, East
Lansing, MI 48824; beers@pa.msu.edu}
 
\and
 
\author{Sean G. Ryan}
 
\affil{Physics Department, The Open University, Walton Hall, Milton Keynes
MK7 6AA, United Kingdom; s.g.ryan@open.ac.uk}

\begin{abstract}

We report high-resolution, high-signal-to-noise, observations of the
extremely metal-poor double-lined spectroscopic binary CS~22876--032.  The
system has a long period : P = 424.7 $\pm$ 0.6 days.  It comprises two main
sequence stars having effective temperatures 6300~K and 5600~K, with a
ratio of secondary to primary mass of 0.89 $\pm$ 0.04.  The metallicity of
the system is [Fe/H] = --3.71 $\pm$ 0.11 $\pm$ 0.12 (random and systematic
errors) -- somewhat higher than previous estimates.

We find [Mg/Fe] = 0.50, typical of values of less extreme halo material.
[Si/Fe], [Ca/Fe], and [Ti/Fe], however, all have significantly lower values,
$\sim$~0.0--0.1, suggesting that the heavier elements might have been
underproduced relative to Mg in the material from which this object formed.
In the context of the hypothesis that the abundance patterns of extremely
metal-poor stars are driven by individual enrichment events and the models of
Woosley and Weaver (1995), the data for CS~22876--032 are consistent with its
having been enriched by a zero-metallicity supernova of mass 30 M$_{\odot}$.

As the most metal-poor near-main-sequence-turnoff star currently known, the
primary of the system has the potential to strongly constrain the primordial
lithium abundance.  We find A(Li) (= log
(N(Li)/N(H))~+~12.00)~=~2.03~$\pm$~0.07, which is consistent with the finding
of Ryan et al. (1999) that for stars of extremely low metallicity A(Li) is a
function of [Fe/H].

\end{abstract}
 

\hspace {-1.0mm}{\it Suggested running title:} \hspace*{3mm} THE MOST
METAL-POOR DWARF, CS~22876--032
 
\section{INTRODUCTION}

In the past two decades a handful of stars with heavy element abundance less
than 1/1000 that of the sun ([Fe/H]\footnote{[Fe/H] = log~(N$_{\rm
Fe}$/N$_{\rm H}$)$_{\rm star}$--log~(N$_{\rm Fe}$/N$_{\rm H}$)$_{\odot}$} $<$
--3.0) have been analyzed at high spectral resolution, with a view to
understanding the production of the elements at the earliest times.  Such
stars have the potential to constrain Big Bang Nucleosynthesis (Ryan, Norris
\& Beers 1999), the nature of the first supernovae (McWilliam et al. 1995;
Ryan, Norris \& Beers 1996; Nakamura et al. 1999), and the manner in which
the ejecta from the first generations were incorporated into subsequent ones
(Ryan et al. 1996; Shigeyama \& Tsujimoto 1998; Tsujimoto \& Shigeyama 1998;
Ikuta \& Arimoto 1999; Tsujimoto, Shigeyama \& Yoshii 1999).  In special
cases, they can even be used to determine the age of the Galaxy (Cowan et
al. 1999).  As relatively simple objects, formed at redshifts $\gtrsim$~4--5,
they nicely complement and constrain abundance results from the more
complicated and less well-understood Lyman--$\alpha$ clouds and Damped
Lyman--$\alpha$ systems currently studied at redshifts z $<$ 3.5 (see eg Ryan
2000).

The most metal-poor objects are generally faint, and studies to date have
been undertaken with somewhat limited signal-to-noise (S/N).  For the six
stars having [Fe/H] $<$ --3.5 and so far analyzed at high resolution (see
Table 6 of Ryan et al. 1996), perusal of the sources suggests that the data
were obtained with a representative S/N per equivalent 0.04 {\AA} pixel of
roughly 30, corresponding to a 3$\sigma$ detection limit for weak lines of
$\sim$~10 m{\AA}.  In some cases this has led to uncertainty as to the
existence of important elements in these objects\footnote{For example, from
Norris et al. (1996) : ``Ba 4554 {\AA} measurements of CD--38$^{\rm o}$245
are rather discordant, ranging from detection at 18 m{\AA} and 19 m{\AA}
(Bessell \& Norris 1984; Primas et al. 1994) to upper limits $<$~10 m{\AA}
and $<$~7 m{\AA} by Molaro \& Castelli (1990) and Peterson \& Carney
(1989).''}.  As one moves to the lowest abundances (or higher temperatures),
the problem of determining reliable abundances can only deteriorate: for
example, in the two near-main-sequence-turnoff dwarfs in the above-mentioned
sample there are fewer than 20 non-Fe lines detected in the preferred
wavelength range (3700--4700 {\AA}), of which fewer than half are stronger
than 20 m{\AA}.
  
To improve the accuracy of the existing abundance analyses we are attempting
to obtain high resolution (R = 40,000) data with S/N = 100 of objects with
[Fe/H] $<$ --3.5.  The purpose of the present work is to present results for
CS~22876--032.  In a subsequent paper we shall present results for
CD--24$^{\rm o}$17504, CD--38$^{\rm o}$245, CS~22172--002, and CS~22885--096.

CS~22876--032 was discovered by Beers, Preston, \& Shectman (1985), and with a
high-resolution abundance determination of [Fe/H] = --4.3 (Molaro \& Castelli
1990; Norris, Peterson, \& Beers 1993) it is currently the most metal-poor
dwarf for which such analysis exists.  The abundance determinations for this
object are, however, compromised to some extent by the fact that it is a
double-lined spectroscopic binary (Nissen 1989) of long but undetermined
period.  We set out to determine the nature of the two components and thus
improve the reliability of the abundances in the system.

CS~22876-032 has the added attraction that it is the main-sequence dwarf
closest in abundance to the material which emerged from the Big Bang. It 
therefore has a potentially important role in constraining the primordial 
abundance of Li, especially if the Li abundance in higher metallicity stars 
has been elevated by Galactic chemical evolution, as we contend elsewhere
(Ryan et al. 1999). However, only after its atmospheric parameters have been
reliably determined will one be in a position to fully realize that potential.

The outline of the paper is as follows. The observational material upon which
our analysis is based is presented in \S~2, while in \S~3 we use the
available material to present a first estimate of the period and orbital
elements of the system.  These results are coupled with the observed {\it
BVRI} colors and available isochrones in \S~4 to determine the atmospheric
parameters T$_{\rm eff}$, log~$g$, [Fe/H], and ${\xi}$ of the components,
together with element ratios [X/Fe].  Finally, in \S~5 we discuss the results
and revisit the question of the Li abundance of the system, and its role in
constraining the primordial value.

\section{OBSERVATIONAL MATERIAL}

\subsection{High-resolution, High-signal-to-noise Spectroscopy}

CS~22876--032 was observed with the University College London coud\'e\
\'echelle spectrograph, Anglo-Australian Telescope combination during
sessions in August 1996, 1997, and 1998.  The instrumental setup and data
reduction techniques were as described by Norris, Ryan, \& Beers (1996), and
will not be discussed in detail here, except in particular instances where
different techniques were dictated by the binarity of the system.  Suffice it
to say that the material covers the wavelength range 3700--4700~{\AA} and was
obtained with resolving power 40000.  The numbers of detected photons in the
spectra were 2700, 5800, and 3300 per 0.04 {\AA} pixel at 4300 {\AA} at the
three epochs, respectively.

A comparison of the spectra in the region of the Ca II K line, registered to
the rest frame of the primary of the system, is presented in Figure~1.  Note
the binary motion of the secondary as seen in the Fe I lines and the
asymmetry of the Ca II line.  In an effort to best use the effective S/N
$\sim$100 of the combined data set, equivalent widths were determined for
each binary component in two ways.  First, since the lines of the two stars
are well separated in the 1996 and 1997 data, the spectra from these epochs
were co-added in the rest frame of each component in turn, and equivalent
widths measured using techniques similar to those described in Norris et
al. (1996).  The second method utilized the relationship between central line
depth and equivalent width.  All three spectra were co-added in the rest
frame of the relevant component, and central line depths measured.  A
quadratic was then fit to the dependence of these central depths on the
equivalent widths which had been determined from the first method.  This
relationship was then used to determine equivalent widths for all lines
having measured central depths.  The final adopted equivalent widths, W$_{\rm
A}$ and W$_{\rm B}$, are the average of the two methods, and are presented in
the final two columns of Table~1\footnote{It may be of interest to some
readers to compare our final equivalent widths with those determined with the
first, more standard, method.  The mean difference between final and first
estimates is less than 1\% for both primary and secondary.  The mean absolute
differences are 4 and 6 \%, respectively.}.  Throughout this work we shall
denote the {\it observed} properties of primary and secondary with subscripts
A and B, respectively, while for the inferred {\it intrinsic} parameters we
shall use subscripts P and S.  Based on the number of photons detected, we
expect the equivalent widths to be accurate to 1--2 m{\AA}.  The first three
columns of the table present wavelength, excitation potential, and log~{$gf$}
values from sources identified in Norris et al. (1996).  For undetected
species, limits of 5--10 m{\AA} are indicated in the table: these are a
little larger than $\sim$3$\sigma$ limits based on photon statistics, and are
commensurate with the detectability of lines on the spectra.  A comparison of
the present results for CS~22876--032 with those of Molaro \& Castelli (1990)
and Norris et al. (1993) is presented in Figure~2, where one sees that the
presently derived equivalent widths are somewhat smaller than reported by
them.  We conjecture that this may have resulted from the continuum having
been placed too high on the lower S/N spectra in these earlier
investigations.

For lines stronger than 10 m{\AA} in both members of the system, Figure~3
compares the ratio of primary to secondary equivalent widths as a function of
primary line strength, wavelength, and excitation potential.  One sees
little, if any, dependence on wavelength and excitation, but an apparent
correlation with line strength.  Note too the much smaller scatter in the
W$_{\rm A}$/W$_{\rm B}$ vs W$_{\rm A}$ diagram, which is more commensurate
with our error estimates, shown in the figure for $\sigma$$_{\rm W}$ = 2
m{\AA}. This may be understood in terms of curve-of-growth effects, to which
we shall return in \S~4.1.

In \S~5.2 we shall introduce spectra taken at longer wavelengths to determine
the strength of the Li I 6707 {\AA} line.  These data permitted us to
determine the strength of the Mg I b lines in both components, which are also
presented in Table 1.  The S/N per 0.04 {\AA} bin at $\lambda$5180{\AA} was
55.

\subsection{Radial Velocity Data}

Table~2 contains the radial velocities for CS~22786--032 which we have been
able to glean from the archives of the Anglo-Australian Observatory
(principally from our own observations), literature material kindly provided
by Drs. P.E.~Nissen and J.A.~Thorburn, and a service observation from the
William Herschel Telescope.  For the AAO and WHT data we measured velocities
using techniques similar to those described by Norris et al. (1996), which we
shall not repeat here and to which we refer the reader.  The information in
the table should be self-explanatory, except perhaps for the errors, which
provide an estimate of combined internal and external uncertainties.

\subsection{{\it BVRI} and Str\"{o}mgren Photometry, and Interstellar Reddening}

Broad-band {\it BVRI} photometry is available for CS~22876--032 from the work
of Norris et al. (1993), following whom we adopt {\it B--V} = 0.395, {\it
V--R} = 0.28, and {\it R--I} = 0.31.  These values have uncertainties of 0.01
mag.  Str\"{o}mgren photometry is available from Schuster, Parrao, \&
Contreras Martinez (1993) and Schuster et al. (1996): {\it b--y} = 0.334,
{\it m}$_{1}$ = 0.036, and {\it c}$_{1}$ = 0.245, with uncertainties
$\sim$0.005.  Following the latter authors we adopt E({\it B--V}) = 0.00,
based on the analysis of their Str\"{o}mgren data.  Beers, Preston, \&
Shectman (1992) report E({\it B--V}) = 0.01 for CS~22876--032 in good
agreement with this value, suggesting that the reddening is well-determined,
with uncertainty 0.01 mag.

\section{ORBITAL ELEMENTS}

The 21 radial velocity measurements are sufficient to constrain considerably
the orbital elements of this system.  In Figure~4a we present the periodogram
of the primary, determined using the procedure of Lafler \& Kinman (1965).
Our best estimate of the period from this method is $\sim$425 days.  Values
corresponding to other less prominent dips in the figure result in the phased
velocity diagram having considerably larger scatter than is seen for this
value.

An initial estimate of the orbital elements for the two components was
obtained by first setting the orbital period fixed at 425 days, and solving
only for the elements of the primary using the method of Russell-Wilsing
(Russell 1902; re-discussed by Binnendijk 1960).  We then obtained estimates
of the orbital elements for both the primary and secondary of CS~22876--032
following the differential corrections procedure of Lehmann-Filh\'es (1894;
re-discussed by Underhill 1966), using code kindly made available to us by
Dr.  R.D. Mathieu (private communication).  The code we employed, SBOP, is an
adaptation of a FORTRAN II code originally listed by Wolfe, Horak, \& Storer
(1967).  Solutions were attempted with an application of equal weights to all
observations, followed by a de-weighting of the two observations made with
the RGO spectrograph, in order to reflect their lower velocity accuracy.  The
elements which were obtained were quite similar in both cases.  Results are
presented in columns (2) and (3) of Table~3.  The observational data, phased
to these elements, together with the model radial velocity curves for primary
and secondary are presented in Figure~4b.  The agreement is quite
satisfactory.

At the suggestion of the referee, we have attempted to obtain orbital
solutions with starting values for the period centered around the less
prominent dips in the periodogram, at P $\sim$316 and P $\sim$500 days,
respectively.  In both instances, SBOP failed to converge to a satisfactory
solution, consistent with our expectations.  As a check on the sensitivity to
the initial guess of the orbital period, we have also sought solutions with
starting values for the period in the range 405 to 450 days.  Outside this
range of initial periods, SBOP again failed to converge to a solution.  For
starting points within the above range, SBOP converged to the same final
period estimate, and the same set of orbital parameters as when we chose a
starting period of 425 days.  We conclude that our final solutions are
stable, and accurate to the limit which the present data allow.  Future
spectroscopic study of CS 22876--032 will allow us to more fully populate the
phase diagram, and hence enable refinement of the derived orbital elements.

The errors on all of the estimated elements are also listed in Table~3.  As
can be seen from inspection of the table, the elements are well constrained
with the exception of the time of periastron passage (T$_{0}$) and the angle
of periastron passage ($\omega$).  This is perhaps to be expected given the
low eccentricity of this system (e $\sim$~0.12) and the relatively sparse
phase coverage of our measurements.  The mass ratio of the system M$_{\rm
S}$/M$_{\rm P}$ = $0.89 \pm 0.04$ provides a useful confirmation of our
attempt at deriving appropriate atmospheric parameters for the two components
by comparison with model isochrones, discussed below. Even though the error in
T$_0$ is larger than we would like, it corresponds to only 0.03 in phase, so
one still obtains useful predictions of optimal observing opportunities when
the components are expected to show the greatest velocity separation.

For heuristic purposes we present in column (4) of the table the changes to
the elements which were obtained when the two most discrepant data points
(which have large observational uncertainties) were removed from the data
set.  While this leads to elements with slightly improved errors, it is
interesting to note that the differences are well within the cited errors in
column (3).

We now consider whether CS~22876--032 is an eclipsing system.  Given the
large separation of this pair of dwarfs the likelihood is not great.  Assuming
solar diameters, we estimate that the chance of eclipse is less than 0.02,
and its duration less than one day.  We also note that while the period of
the system is established to $\sim$~0.6 day, the zero point of the ephemeris
is known to only 14 days, which will make the search for eclipse more
difficult.  That said, the additional insight to be obtained from eclipse
information would seem to warrant such an effort.

\section{ATMOSPHERIC PARAMETERS}

\subsection{T$_{\rm eff}$, log~$g$, [Fe/H]}

One may use the observed {\it BVRI} and Str\"{o}mgren colors together
with stellar evolutionary isochrones to constrain the atmospheric
parameters of the components of the system.

With ({\it B--V})$_{0}$ = 0.395, CS~22876--032 could in principle comprise a
system having either two objects below the halo main-sequence turnoff, or a
system with one or both objects above it.  The Str\"{o}mgren c$_{1}$ value of
0.245, however, which is typical of metal-poor main sequence dwarfs (see
Schuster et al. 1996), clearly requires that both objects lie below the
turnoff.  If, for example, one forms composite colors from pairs of models on
the VandenBerg \& Bell (1985) isochrones with Y = 0.20, Z = 0.0001, and age =
14 Gyr, one finds that with {\it B--V} $\sim$~0.40, c$_{1}$ $\sim$~0.35 for
those containing at least one star above the turnoff, and c$_{1}$ $\sim$~0.25
for those with both below it.  In what follows, therefore, we shall assume
that both components of CS~22876--032 lie below the main sequence turnoff.

For convenience we illustrate our method using the Revised Yale Isochrones
(Green, Demarque, \& King 1987 (RYI)), for Y = 0.20, Z = 0.00001
([Fe/H] $\sim$~--2.3), and age = 13 Gyr, and later comment on its sensitivity
to these choices.  Figure~5a,b show the locus of isochrone pairs of {\it
B--V} and of T$_{\rm eff}$ values which together combine to produce the value
({\it B--V})$_{0}$ = 0.395 observed for CS~22876--032.  One sees that only a
narrow range of permitted color and T$_{\rm eff}$ exists for the primary, and
one can thus constrain the parameters for this component quite well.  The
situation is not so favorable for the secondary, but once the primary is
specified one can constrain the secondary by requiring that both stars have
the same metallicity.  An iterative procedure, which converged quickly, is
described below.

For any isochrone pair one may also determine the equivalent width correction
factors one should apply to observed line strengths to obtain intrinsic
values.  (The corrected equivalent widths are obtained by multiplying the
observed ones by these factors.)  The correction factors f$_1$ and f$_2$ are
the reciprocals of the fractional flux contributions of the two components,
so 1/f$_1$ + 1/f$_2$ = 1.0, and the primary:secondary luminosity ratio is
simply f$_2$/f$_1$. The values are inferred directly from the isochrones.
Strictly, the (f$_1$, f$_2$) values are wavelength dependent, but in practice
a single pair suffices throughout the blue-violet spectral region.  We note
for completeness that the values adopted here are determined at the effective
wavelength of the B band.  Fluxes are sufficiently different in the red
spectral region, however, that red-specific ratios must be utilized for the
analysis of Li~6707~\AA (see below).  In Figure~5c we show the locus of pairs
of correction factors applicable in the blue-violet region.
 
Similar diagrams may be constructed for {\it V--R} and {\it R--I}.  Given the
observational data of Table~1, one is now in a position to determine
abundances for model pairs which reproduce the observed colors. We refer the
reader to Ryan, Norris, \& Beers (1996) for the details of our model
atmosphere abundance determination technique, which are quite standard and
which we shall not repeat here, except to note that we employ Bell (1983)
models in our analysis.  Figure~6 shows the dependence of [Fe/H] on T$_{\rm
eff}$ for several pairs of permitted primary, secondary pairs, determined
when using atmospheric parameters based on {\it B--V}.  By adopting the
requirement that both components have the same value of [Fe/H] (to within
0.01 dex) we then remove the degeneracy of the procedure.  Similar diagrams
were constructed using {\it V--R} and {\it R--I} rather than {\it B--V} to
constrain the atmospheric parameters.  Table~4 contains the results, where
column (1) presents the observed composite color, columns (2), (3), (4), and
(5) contain the mass, T$_{\rm eff}$, color, and equivalent width correction
factor for the primary, respectively, and columns (6)--(9) give the same
parameters for the secondary.  Assigning equal weight to the values for each
of the three cases we find that the system has [Fe/H] = --3.71 and that the
masses are 0.83 and 0.73 M$_{\odot}$.  The average mass ratio is 0.88, in
pleasing accord with the analysis of orbital parameters in \S~3.  Finally, in
Table~5 we present our adopted atmospheric parameters for the components of
the system.

In Figure~3 we showed that the ratio of observed equivalent widths, W$_{\rm
A}$/W$_{\rm B}$, increases with line strength. The equivalent width
correction factors derived in the decomposition of the binary, f$_{\rm
4300,P}$ and f$_{\rm 4300,S}$, are listed in Table~4 and aid in explaining
the observed variation of the ratio of the observed equivalent
widths. Specifically, the large value of f$_{\rm 4300,S}$ indicates that the
{\em apparently} weak lines in the secondary (W$_{\rm B}$ in the range
8--24~m{\AA}) are in reality stronger by a factor $\sim$~4.6 in the
uncontaminated spectrum, and thus occupy the moderately strong range
37~$\le$~W$_{\rm S}$~$\le$~110~m{\AA}, partially on the flat part of the
curve of growth.  The primary, on the other hand, has only a factor 1.28
difference between its observed (W$_{\rm A}$) and intrinsic (W$_{\rm P}$)
values, confirming that most of its lines do indeed lie on the linear portion
of the curve of growth. The increase in the ratio W$_{\rm A}$/W$_{\rm B}$ is
therefore understood as the progressive saturation of the secondary's lines
with increasing line strength, while those of the primary are not so
affected.

We illustrate this in Figure~7, which compares the observations with
theoretical line ratios computed for two stars having the adopted parameter
set. The theoretical curve reproduces the observed behavior quite well,
confirming the origin of this effect.  The reader might comment that the
agreement is less than satisfactory for $W_{\rm A}$ $>$ 40 m{\AA}, and we
would have to concur.  We note, however, that it is in this regime that the
lines of the secondary have $W_{\rm S}$ $\sim$~100--140 m{\AA} and
experience strong saturation effects.  It is not unlikely that uncertainties
in the relatively poorly-defined adopted microturbulence of the secondary and
the treatment of van der Waals broadening (see e.g. Ryan 1998) might lead to
significant error in the theoretical strengths of these lines.

\subsection{Error Budget}

How robust are the abundances and atmospheric parameters in Tables~4 and 5?
The answer is somewhat complicated because of the interplay of the two
components of the binary.  We discuss random and systematic errors, including
the impact of one star upon the other, and summarize the situation in
Table~6.

First consider the uncertainties in the adopted atmospheric parameters.
Errors in color-effective temperature transformations of $\sim$~100~K are
(unfortunately) common in the analyses of metal-poor stars.  If the two
components of the system were of comparable spectral type, both would be
affected similarly by this error source.  Since, however, their temperatures
differ by 700~K, independent errors of 100~K are allocated to each.
Additionally, errors in photometry and reddening are 0.01~mag, each yielding
an additional uncertainty of 40--60~K. The adopted error in each star's
T$_{\rm eff}$, summing in quadrature, is therefore 130~K.  We note that this
is consistent with the range of temperatures derived in Table~4 for the three
photometric colors.  The error in T$_{\rm eff}$ for the secondary, however,
has almost no impact on the derived [Fe/H] of the system, as can be seen in
Figure~6, since an incorrect temperature will be compensated for by a change
in the derived f$_{\rm S}$ factor.  We assign 0.10~dex and 0.03~dex as the
error in [Fe/H] induced by the primary and secondary stars' temperature
uncertainties, respectively.  We adopt the sensitivity to errors in
microturbulence and surface gravity from the calculations, but apply only
those due to the primary, again because of the weak sensitivity to the
secondary's parameters.

Taking all atmospheric parameter uncertainties into account the quadratic sum
of errors in [Fe/H] is 0.11~dex.

Errors in the parameters of the primary and secondary also change the
inferred mass ratio of the system.  There is little latitude for the mass of
the primary, as indicated by the small range of T$_{\rm eff}$ and f$_{\rm P}$
values in Figure~6.  However, the temperature and mass inferred for the
secondary will vary if the primary is revised.  Changes in the primary
corresponding to $\Delta$[Fe/H]~=~0.11~dex (from above) would induce a change
in the secondary corresponding to 0.02 in the mass ratio.  The small
sensitivity of the mass ratio to the decomposition of the system is confirmed
in Table~4, where the range of values is 0.015.

Additional errors are associated with the adoption of a particular model
atmosphere grid, which can lead to differences of 0.10~dex (Ryan et al. 1996,
\S~3.2), and in the isochrones used in the decomposition of the binary.  The
latter effect was assessed by changing the adopted RYI isochrone and viewing
the impact on the derived parameters of the binary.  Reasonable changes in
helium, heavy element abundance, and age, of $\Delta$Y~=~0.05,
$\Delta$log~Z~=~0.5, and $\Delta$Age~=~2 Gyr, respectively, lead to changes
in the abundance of the primary $\Delta$[Fe/H]~$<$~0.04.  The corresponding
change in the mass ratio is also small, $\simeq$~0.02.

We also repeated the temperature and abundance determination exercise at Z =
0.0001 for the isochrones of the RYI, of Bergbusch \& VandenBerg (1992) and
those of the Padua group available in 1999 October at
http://dns2.pd.astro.it/.  Using {\it B--V} to constrain the temperature we
obtained [Fe/H]$_{\rm P}$ = --3.52. --3.51, and --3.38, respectively.  The
difference between the value obtained by adoption of the Padua and the others
is driven essentially by a difference in their adopted {\it B--V}, T$_{\rm
eff}$ transformation, the role of which has already been considered above.
We conclude, therefore, that our abundance determination for CS~22876--032 is
insensitive to the adoption of the Revised Yale Isochrones.

The summary of the error budget is shown in Table~6, where column (1) lists
the error source, column (2) the representative error in the parameter, and
column (3) the error in the abundance of the primary (and hence of the
system) or M$_{\rm S}$/M$_{\rm P}$, as appropriate.  Given these errors we
conclude that the abundance of CS~22876--032 is --3.71 $\pm$ 0.11 $\pm$ 0.12,
where the first and second uncertainties refer to random and systematic
errors, respectively.

It is of interest to compare this value with the significantly lower
abundances reported by Molaro \& Castelli (1990) and Norris et al. (1993),
who found [Fe/H] = --4.29 and --4.31, respectively.  For illustrative
purposes we show in Table~7 abundance changes which result from various
sources between the present value and that of Norris et al. (1993).  In all
but two cases the parameters and assumptions made in the present work lead to
a higher abundance, accumulating to $\sim$~+0.4 when taken one at a time ---
not too much below the reported difference of +0.6.  Although CS~22876-032 is
still the most metal-poor dwarf known, our current best estimate of its
metallicity is considerably less extreme than the first measurements.  

We complete the discussion by noting that in our high-resolution studies and
those of McWilliam et al. (1995) no halo dwarf {\it or} giant is known with
[Fe/H]~$<~-4.0$.  The low-resolution surveys of Beers and his co-workers also
strongly suggest that the metallicity distribution of the Galactic halo cuts
off near this value (e.g. Beers et al. 1998; Norris 1999).

\subsection{Relative Abundances, [X/Fe]}

Using the techniques described in some detail in Ryan et al. (1996) we next
computed the abundances of the other elements.  Results for the two
components are presented in Table~8, where the column headers should be
self-evident except perhaps for column (2), which gives details of the
features used in the abundance determination.  At the metal weakness of
CS~22876--032, one has in most cases only one or two lines from which to
determine the abundance.  If only one feature is available we give its
wavelength in column (2); otherwise the number of lines involved is
tabulated.  The standard errors given in columns (5) and (7) were determined
using techniques described by Ryan et al. (1996, \S~4.3). In brief, they
represent the quadratic addition of abundance errors due to uncertainties in
the atmospheric parameters and the measurement of line strengths.  The reader
should note that our knowledge of the abundances of elements, other than Fe,
observed in CS~22876--032 is based on only eight atomic lines.

Table~8 also contains abundance limits for a number of elements which are
more stringent than previously available, given the higher S/N of the present
investigation.  We note that in most cases limits are given only for the
primary of the system, since they are not very useful for the secondary.  

The present data permit us to revisit the Mn abundance of CS~22876-032.  The
contrast between the previous Mn~I~4030~\AA\ measurement of $W$~=~20~m\AA\
(Molaro \& Castelli 1990; S/N $\simeq$ 55--100 at R $\simeq$ 20000), and ours
at $W_{\rm A}~<~5$~m\AA (S/N $\simeq$ 100 at R $\simeq$ 40000) highlights yet
again (if such a comment is necessary) the need for high S/N and resolving
power to study weak lines. In the absence of a definitive test, we note that
the new measurement leads to a Mn abundance which more closely resembles that
of other stars at this [Fe/H].

With one exception the abundances determined for the two components of
CS~22876--032 are in excellent agreement, consistent with the atmospheric
parameters of both objects having been well determined.  The exception is Si
for which the relative abundances disagree by 0.50, a difference formally
significant at the 1.9$\sigma$ level.  In comparison, the mean absolute
difference for the four other elements observed in both components is 0.05
dex.  Although a 1.9$\sigma$ difference is not highly significant, we have
scrutinized all of our spectra of this line.  We find no reason beyond the
formal errors for mistrusting this measurement.  However, we remain also at a
loss to explain the different Si abundances in the two components.

\section{DISCUSSION}

\subsection{Heavy Element Abundances}

Evidence has mounted over the past decade that below [Fe/H]~=~$-3.0$,
Galactic chemical enrichment was a rather patchy business (e.g. Ryan, Norris
\& Bessell 1991; McWilliam et al. 1995; Ryan et al. 1996), leading to models
in which such metal-poor objects are formed in the interaction of supernova
remnants with material in their immediate vicinity (Ryan et al. 1996;
Tsujimoto et al. 1999). In such models, one expects the abundance patterns of
the most metal-poor stars to be representative of individual supernova,
rather than of a time-averaged ensemble.  Hence, high-resolution
spectroscopic analyses of extremely metal-poor stars provide a powerful probe
of the nature of the first supernovae in the Milky Way.

Given the paucity of lines in CS~22876--032 we have only Al and the
$\alpha$--elements to discuss.  In the context of star-to-star scatter we
shall defer further consideration until we present results on the other four
extremely metal-poor stars for which we have data, and which we mentioned in
\S~1 -- except to note here that the value [Al/Fe] = --0.44 for CS~22876--032
lies within $\sim$~0.20 of the mean line determined for the sample of
metal-poor stars discussed in the investigation of Ryan et al. (1996,
Figure~3).

In the spirit of associating the yields of particular supernovae models with
the observed abundances in extremely metal-poor stars, we conclude by
discussing the $\alpha$--elements.  For Mg, which is produced mainly during
hydrostatic carbon and neon burning we find [Mg/Fe] = 0.50.  For Si and Ca we
find a weighted mean value, which we designate [$\langle$Si,Ca$\rangle$/Fe],
of --0.05 (the data were weighted by the inverse square of their errors).
These elements are produced by a combination of hydrostatic oxygen shell
burning and explosive oxygen burning, which varies from star to star (Weaver
\& Woosley 1993). As discussed by Woosley \& Weaver (1995), the predicted
yields of Si and Ca are expected to vary relative to Mg, depending
sensitively on several physical effects.  (These include among others the
treatment of convection, the density structure near the iron core, the
location of the mass cut, and the amount of material which falls back in the
explosion.)  For CS~22876--032 we have [$\langle$Si,Ca$\rangle$/Mg] = --0.55.
It is then interesting to consider the production factors of the
zero-heavy-element model supernovae of Woosley \& Weaver (1995, Table 17).
For those models which produce significant amounts of Mg, Si, and Ca we find
the results presented in Figure~8.  If one were to seek to interpret the data
for CS~22876--032 in terms of enrichment from a single zero-heavy-element
supernova one could identify it, quite reasonably, with the Woosley \& Weaver
(1995) 30 M$_{\odot}$ model Z30B.

Titanium is produced under more extreme conditions than Si and Ca, as
discussed by Woosley \& Weaver (1995).  In CS~22876--032, as for Si and Ca,
[Ti/Fe] is low relative to [Mg/Fe]: we find [Ti/Mg] = --0.39.  This is also
best reproduced by Woosley \& Weaver model Z30B.

Is our result of [Al/Mg] = --0.94 $\pm$ 0.17 consistent with the
hypothesis?  Before comparison with theory one must correct the present LTE
abundance for Al for non-LTE effects.  According to Baum\"{u}ller \& Gehren
(1997), LTE abundances for Al based on the line we have used underestimate
the abundance by $\sim$~0.65 dex at [Fe/H] $\sim$~--2.5$-$--3.0. Adopting this
correction for CS~22876--032, one therefore has [Al/Mg] $\sim$~--0.3 $\pm$
0.2. For the above-mentioned models of Woosley \& Weaver (1995), in
comparison, one finds [Al/Mg] in the range $\sim$~--0.65$-$--0.35.  That is,
the models do not strongly constrain the situation, and within the
observational errors one has reasonable agreement between theory and
observation.

\subsection{Lithium Abundance}

Given the improved atmospheric parameters for the components of
CS~22876--032, one may revisit the question of its lithium abundance.  The
best currently available observational material is that of Thorburn \& Beers
(1993).  They obtained data with S/N = 150 and reported W$_{\rm A}$(Li I
6707{\AA}) = 11 $\pm$ 1.3 m{\AA} at JD 2448850.8, when the lines of the two
components were well separated.  They did not detect the Li line in the
secondary.  Norris et al. (1994) have also observed this object and report
W$_{\rm A}$(Li I 6707{\AA}) = 15 $\pm$ 2.1 m{\AA} from a spectrum with S/N =
70, obtained at JD 2448490.5 when the lines of the primary and secondary
would have been well separated, by some 20 km~s$^{-1}$.

As part of our continuing interest in the lithium problem we attempted to
remeasure the equivalent width during an observing run on the
Anglo-Australian Telescope in 1999 September, using techniques similar to
those described by Norris et al. (1994) and Ryan, Norris, \& Beers (1999).
From a spectrum having S/N = 70 per 0.04 {\AA} increment we obtain W$_{\rm
A}$(Li I 6707{\AA}) = 14.4 $\pm$ 2.3 m{\AA}.  At the time of observation the
lines of the primary and secondary were separated by 21 km s$^{-1}$, and the
line in the secondary was below our threshold of detectability, which we
estimate to be 7 m{\AA} (3$\sigma$).

We seek to compare these data with the recent accurate and homogeneous
observational material of Ryan et al. (1999).  We first ask if there are
systematic equivalent width errors between the various authors.  For 16 stars
in common between Ryan et al. (1999) and Thorburn (1994) (who used the same
equipment and techniques as those of Thorburn \& Beers) one finds
$\langle$W$_{\rm RNB}$--W$_{\rm T}$$\rangle$ = --1.9 $\pm$ 1.1 (s.e.) m{\AA},
while for 5 stars in common between Ryan et al. and Norris et al. (1994) one
finds $\langle$W$_{\rm RNB}$--W$_{\rm NRS}$$\rangle$ = +2.3 $\pm$ 1.7 (s.e.)
m{\AA}.  We do not regard these differences as statistically significant, and
make no correction to the reported equivalent widths.  We then adopt W$_{\rm
A}$(Li~I 6707{\AA}) = 12.5 $\pm$ 1.4 m{\AA} as the best currently available
value for CS~22876--032, having weighted the data by the inverse square of
their estimated errors.  (The error is based on small sample statistics
following Keeping (1962) without cognisance of the weights.)  At
$\lambda$6707{\AA} our model shows that to obtain the line strength intrinsic
to the primary this should be corrected by the factor f$_{6707,\rm P}$ =
1.35\footnote{Based on {\it B-R} results from the RYI and on model atmosphere
synthetic spectra, which give consistent results.}, which leads to W$_{\rm
P}$(Li~I 6707{\AA}) = 16.9 $\pm$ 1.9 m{\AA}.

The recent study of Li by Ryan et al. (1999) used an effective temperature
scale different from the RYI one that accompanies the isochrones used in the
decomposition of the binary. It would be inappropriate to compare the Li
abundance of CS~22876-032 computed using the RYI temperature scale, with a
set of data based on an entirely different one. For this reason, we use the
decomposed {\it colors} of the binary, as given in Table~4, to infer the
temperature on the same scale as our 1999 Li work.  We obtain T$_{\rm
eff}$~=~6223~$\pm$~38~K.  The inferred lithium abundance\footnote{The Li
abundance was calculated using exactly the same formalism as in Ryan et
al. (1999), i.e. computing synthetic spectra for Bell (1983) models using
four $^7$Li components, and measuring the equivalent width of the synthesized
line.}  A$_{\rm P}$(Li) (= log (N$_{\rm P}$(Li)/N$_{\rm
P}$(H))~+~12.00)~=~2.03~$\pm$~0.07\footnote{The perceptive reader will note
that in this work we have combined the higher S/N data of Thorburn \& Beers
(1993; S/N = 150) with our lower S/N data (two spectra each with S/N = 70),
whereas in a recent work (Ryan et al. 1999) on the spread about the plateau
we advocated the importance of homogeneity. We would not advocate the
coaddition of data were we studying the spread here, since the statistic used
to describe spread -- the standard deviation -- is acutely sensitive to
errors, which enter as squared terms. However, the current comparison is with
the mean abundance of the stars, and the mean is less affected than the
standard deviation by divergent phenomena. Consequently we have chosen to use
the superior-S/N data of Thorburn \& Beers as well. We note that if we
instead adopted only the less accurate (S/N-limited) value of $W_{\rm A}$ =
14.4$\pm$2.3~m{\AA} reported in the present work, the Li abundance would be
0.06~dex higher and the uncertainties larger at $A_{\rm P}$(Li) = 2.09 $\pm$
0.08 (1$\sigma$). In this case the star moves further above the trend, but on
account of its much larger error bars, has little impact on the slope.}.
(The uncertainty in the quadratic addition of errors corresponding to
$\Delta$f$_{\rm 6707,P}$~=~0.01, $\Delta$T$_{\rm eff}$~=~38~K, and
$\Delta$W$_{\rm P}$(Li I 6707\AA)~=~1.9~m\AA.  We recall that here `P' by
convention refers to `Primary' and not necessarily `Primordial'.)

The non-detection of the secondary by Thorburn \& Beers (1993) and the
present work is also important.  If this component has T$_{\rm eff}$ = 5600~K
and the same Li abundance as the primary, one would expect an intrinsic line
strength W$_{\rm S}$(Li~I 6707{\AA}) = 45 m{\AA}.  Applying a model
correction factor f$_{6707, \rm S}$ = 3.9 (applicable at $\lambda$6707 {\AA}
for the secondary), the observed line strength would then be 12 m{\AA}.  The
most likely explanation for the non-detection of Li in the secondary is that
with T$_{\rm eff}$ = 5600~K it lies at the cool edge of the Spite Plateau and
has experienced modest depletion to below its primordial value.  The
uncertainty in the temperature of the secondary is also greater than that of
the primary, as is evident from Figure~6, and could explain part of the
problem.

From their investigation of an unbiased sample of 23
near-main-sequence-turnoff stars having 6050~K $<$ T$_{\rm eff}$ $<$ 6350~K
and --3.6 $<$ [Fe/H] $<$ --2.3 Ryan et al. (1999) reported a dependence of
A(Li) on [Fe/H].  Their "best fit" was A(Li) = 2.447($\pm$0.066) +
0.118($\pm$0.023)$\times$[Fe/H].  They argued that the most likely
explanation of the positive trend with metallicity was that the primordial
lithium abundance had been increased by early Galactic chemical enrichment
involving the interaction of cosmic rays with the interstellar medium (see
also Ryan et al. (2000)).  Their material and its "best fit" trend, together
with the present result for CS~22876--032, are presented in Figure~9.  It is
clear that the data for the latter object, which is currently the lowest
metallicity dwarf known, are completely consistent with the trend
reported by Ryan et al. and support their hypothesis.  Equally clearly, as
noted by those authors, more data are required at [Fe/H] $<$ --3.5 to place
their result, and its implications for the primordial Li abundance, on a
firmer basis.

We also note that to infer the primordial abundance from the observed one,
a range of potentially systematic as well as random uncertainties needs to be
taken into consideration. Furthermore, the computations of the primordial Li
abundance as a function of baryon-to-photon ratio, $\eta$, or equivalently
$\Omega_{\rm B}$, are themselves subject to significant uncertainties. The
impact of these issues is discussed more fully elsewhere (Ryan et al. 2000).

\acknowledgements

We are grateful to the Director and staff of the Anglo-Australian
Observatory, and the Australian Time Allocation Committee for providing the
observational facilities used in this study.  We are likewise grateful to Dr
D. Folha for obtaining the WHT spectrum as a Service Observation.  It is a
pleasure to thank Drs. R.D. Mathieu, P.E. Nissen, and J.A. Thorburn, for
their generous assistance.  J.E.N. gratefully acknowledges the support of the
Department of Astronomy, University of Texas at Austin during the preparation
of the manuscript, and T.C.B. acknowledges support of NSF grants AST
92--22326, AST 95--29454, and INT 94--17547.  We also gratefully acknowledge
the thorough report from Dr. P. Bonifacio, which led to a number of
improvements to the text.

\clearpage
\begin{center}
\begin{deluxetable}{ccccc}
\tablecaption{EQUIVALENT WIDTHS FOR CS~22876--032A,B}
\tablehead{
\colhead{$\lambda$}& {$\chi$}& {log~$gf$}& {W$_{\rm A}$} & {W$_{\rm B}$}\\ 
        {(\AA)}& {($\rm eV$)} & & {(m\AA)} & {(m\AA)}
\nl 
        {(1)} &{(2)} &{(3)} & {(4)} &{(5)} 
 }
\startdata
 Mg I  &     &      &     &       \nl 
3829.35& 2.71& --0.48&   57&     29\nl
5172.70& 2.71& --0.38&   51&     24\nl
5183.62& 2.72& --0.16&   59&     29\nl
 Al I  &     &      &     &       \nl  
3944.01& 0.00& --0.64&   11&     11\nl
3961.52& 0.00& --0.34&   15&     11\nl
 Si I  &     &      &     &       \nl   
3905.52& 1.91& --1.09&   15&     17\nl
 Ca I  &     &      &     &       \nl   
4226.73& 0.00& +0.24&   34&     18\nl
 Sc II &     &      &     &       \nl 
4246.82& 0.32& +0.32&   $<$5&     $<$5\nl
 Ti II &     &      &     &       \nl 
3759.30& 0.61& +0.20&   28&     14\nl
3761.32& 0.57& +0.10&   21&      9\nl
 Cr I  &     &      &     &       \nl 
4254.33& 0.00& --0.11&   $<$5&     $<$5\nl
 Mn I  &     &      &     &       \nl 
4030.75& 0.00& --0.62&   $<$5&     $<$5\nl
 Fe I  &     &      &     &       \nl   
3727.63& 0.96& --0.62&   24&     12\nl
3743.37& 0.99& --0.78&   27&     $<$10\nl
3758.24& 0.96& --0.02&   44&     17\nl
3763.80& 0.99& --0.23&   33&     19\nl
3787.88& 1.01& --0.85&   19&     $<$10\nl
3812.96& 0.96& --1.03&   13&     13:\nl
3815.84& 1.48& +0.24&   33&     20\nl
3820.43& 0.86& +0.14&   57&     26\nl
3825.88& 0.92& --0.03&   47&     23\nl
3827.82& 1.56& +0.08&   28&     19\nl
3849.97& 1.01& --0.87&   17&      7\nl
3856.37& 0.05& --1.28&   38&     17\nl
3859.91& 0.00& --0.70&   62&     22\nl
3865.52& 1.01& --0.97&   14&     12:\nl
3872.50& 0.99& --0.91&   19&     15\nl 
3878.02& 0.96& --0.91&   14&     $<$7\nl 
3878.57& 0.09& --1.36&   33&     18\nl 
3899.71& 0.09& --1.52&   28&     18\nl 
3920.26& 0.12& --1.74&   21&     15\nl 
3922.91& 0.05& --1.64&   24&     15\nl 
3927.92& 0.11& --1.52&   24&     16\nl 
3930.30& 0.09& --1.49&   28&     17\nl 
4005.24& 1.56& --0.60&   12&     10:\nl
4045.81& 1.48& +0.28&   41&     21\nl 
4063.59& 1.56& +0.06&   30&     17\nl 
4071.74& 1.61& --0.02&   24&     16\nl 
4132.06& 1.61& --0.68&   10&     $<$5\nl 
4143.87& 1.56& --0.20&   13&      9:\nl
4202.03& 1.48& --0.70&   10&      9:\nl
4250.79& 1.56& --0.71&    8&     $<$5\nl 
4260.47& 2.40& +0.11&    8&      $<$5\nl 
4271.76& 2.45& --0.34&   28&     12\nl 
4307.90& 1.56& --0.07&   25&     15\nl   
4325.76& 1.61& +0.01&   25&     16\nl 
4383.54& 1.48& +0.20&   37&     16\nl 
4404.75& 1.56& --0.13&   23&     15\nl  
4415.12& 1.61& --0.62&   11&      7:\nl
 Co I  &     &      &     &       \nl 
3873.11& 0.43& --0.66&    $<$7&     $<$7\nl
 Ni I  &     &      &     &       \nl 
3858.29& 0.42& --0.95&   $<$7&     $<$7\nl
 Sr II &     &      &     &       \nl 
4077.71& 0.00& +0.15&   $<$5&     $<$5\nl
 Ba II &     &      &     &       \nl 
4554.03& 0.00& +0.16&   $<$5&     $<$5
\enddata 
\end{deluxetable}
\end{center}

\clearpage
\begin{center}
\begin{deluxetable}{lcrrrrl}
\tablecaption{RADIAL VELOCITIES FOR CS~22876--032A,B}
\tablehead{
\colhead{Date} & {JD} &{V$_{\rm A}$} &{$\Delta$V$_{\rm A}$} & {V$_{\rm B}$}& {$\Delta$V$_{\rm B}$} & {Source}
\nl 
        {(1)} &{(2)} &{(3)} & {(4)} &{(5)} &{(6)} &{(7)}
}
\startdata
1985 Sep 6  &2446315.1   & --83.9 & 2.0 &   ... & ... & AAT, RGO\tablenotemark{a}     \nl       
1985 Dec 16 &2446416.0   & --96.8 & 2.0 &   ... & ... & AAT, RGO\tablenotemark{a}     \nl      
1989 Sep 13 &2447783.2   &--107.8 & 1.0 & --74.8& 1.0 & AAT, UCLES\tablenotemark{b}   \nl
1989 Oct 16 &2447815.7   &--103.1 & 1.0 & --81.2& 1.0 & P.E. Nissen\tablenotemark{c}  \nl   
1989 Dec 6  &2447867.0   & --93.3 & 1.0 &   ... & ... & AAT, UCLES\tablenotemark{b}   \nl   
1990 Sep 27 &2448162.1   &--109.6 & 1.0 & --76.2& 1.0 & AAT, UCLES\tablenotemark{b}   \nl    
1991 Aug 22 &2448490.5   & --87.0 & 2.0 &   ... & ... & AAT, UCLES\tablenotemark{b}   \nl        
1992 Aug 17 &2448851.8   & --78.6 & 1.0 &--107.6& 1.0 & J.A. Thorburn\tablenotemark{d}\nl   
1996 Aug 7  &2450303.1   &--110.2 & 1.0 & --76.2& 1.0 & AAT, UCLES\tablenotemark{b}   \nl   
1997 Aug 23 &2450683.8   &--104.2 & 1.0 & --81.3& 1.0 & AAT, UCLES\tablenotemark{b}   \nl    
1998 Aug 12 &2451038.3   & --87.8 & 1.0 &--101.1& 1.0 & AAT, UCLES\tablenotemark{b}   \nl   
1999 Jul 29 &2451388.7   & --77.0 & 2.0 &   ... & ... & WHT, UES\tablenotemark{e}     \nl   
1999 Sep 23 &2451444.6   & --84.3 & 1.0 &--105.5& 1.0 & AAT, UCLES\tablenotemark{b}   \nl 

\enddata 
\tablenotetext{a}{AAT, RGO spectrograph combination}
\tablenotetext{b}{AAT, University College London coud\'e\ \'echelle spectrograph combination}
\tablenotetext{c}{Private communication from P.E. Nissen.  See Nissen (1989)}
\tablenotetext{d}{Private communication from J.A. Thorburn.  See Thorburn \& Beers (1993)}
\tablenotetext{e}{WHT, Utrecht \'echelle spectrograph combination; this closely matches the AAT, UCLES setup}
\end{deluxetable}
\end{center}

\clearpage
\begin{center}
\begin{deluxetable}{lrrrr}
\tablecaption{ORBITAL ELEMENTS OF CS~22876--032}
\tablehead{
\colhead{Parameter}                           &  {Value}           & {s.e}  & {$\Delta$\tablenotemark{a}}    \nl 
                             {(1)}            &{(2)}               &{(3)}   &{(4)}        \nl
}
\startdata
Orbital period (days)                         &       424.71     &    0.60 & +0.18   \nl
T$_{0}$  (JD--2400000)                         &     48576.37     &   13.51 & +2.67   \nl
Eccentricity                                  &         0.12     &    0.03 & --0.01  \nl
$\omega$ (degrees)                            &       144.96     &   12.40 & +2.95   \nl
Center-of-mass  radial velocity (km s$^{-1})$ &     $-$93.36     &    0.28 & --0.02  \nl
&&\nl
Primary:&\nl
Radial vel. semi-amplitude (K$_{\rm P}$) (km s$^{-1}$) &15.13          &  0.51 & +0.03 \nl
Projected semi-major axis (a$_{\rm P}$sin(i)) (km)  & 8.77 10$^{7}$ &  3.00 10$^{6}$ & 0.03 10$^{7}$ \nl
M$_{\rm P}$sin$^{3}$(i) (M$_{\odot}$)                      & 0.76          &  0.04 & 0.01  \nl
&&\nl
Secondary:&\nl
Radial vel. semi-amplitude (K$_{\rm S}$) (km s$^{-1}$) &17.06          &  0.56 & 0.08 \nl
Projected semi-major axis (a$_{\rm S}$sin(i)) (km)  & 9.89 10$^{7}$ &  3.25 10$^{6}$ & 0.06 10$^{7}$ \nl
M$_{\rm S}$sin$^{3}$(i) M$_{\odot}$                        & 0.68          &  0.04 & 0.01 \nl
&&\nl 
M$_{\rm S}$/M$_{\rm P}$                        &      0.89               &   0.04  & 0.00 \nl

\enddata 
\tablenotetext{a}{Element change when two most discrepant points are removed from the sample.}
\end{deluxetable}
\end{center}
  
\clearpage
\begin{center}
\begin{deluxetable}{lcccccccccc}
\tablecaption{PHYSICAL PARAMETERS FOR CS~22876--032}
\tablehead{
\colhead{Color$_{\rm A+B}$} &{M$_{\rm P}$} &{T$_{\rm P}$} &{Color$_{\rm P}$} &{f$_{\rm 4300,P}$\tablenotemark{a}}& {M$_{\rm S}$} &{T$_{\rm S}$} &{Color$_{\rm S}$} &{f$_{\rm 4300,S}$\tablenotemark{a}}& {[Fe/H]} & {M$_{\rm S}$/M$_{\rm P}$}
\nl 
        {(1)} &{(2)} &{(3)} & {(4)} &{(5)} &{(6)} &{(7)} & {(8)} & {(9)} & {(10)} & {(11)}
 }

\startdata

Individual:\nl
~~{\it B--V} = 0.395& 0.841& 6425& 0.360& 1.28& 0.743& 5700& 0.515& 4.6& --3.60& 0.883 \nl
~~{\it V--R} = 0.280& 0.825& 6250& 0.258& 1.29& 0.723& 5575& 0.345& 4.5& --3.75& 0.876 \nl
~~{\it R--I} = 0.310& 0.819& 6225& 0.287& 1.26& 0.711& 5500& 0.375& 4.9& --3.78& 0.868 \nl

Mean:\nl    
~~            & 0.828& 6300&      & 1.28& 0.726& 5592&      & 4.7& --3.71& 0.876 \nl
\enddata 
\tablenotetext{a}{Equivalent width correction factor at $\lambda$4300 {\AA}.}\nl

\end{deluxetable}
\end{center}

\clearpage
\begin{center}
\begin{deluxetable}{cccccc}
\tablecaption{ADOPTED ATMOSPHERIC PARAMETERS FOR CS~22876--032}
\tablehead{
\colhead{Component} &{T$_{\rm eff}$} &{log~$g$} &{[Fe/H]} &{$\xi$\tablenotemark{a}} &{f$_{\rm 4300}$}
\nl 
        {(1)}       &{(2)}           &{(3)}     & {(4)}   &{(5)}   &{(6)}
}

\startdata
Primary             & 6300           &4.5       &--3.71    &1.4     &1.28\nl
Secondary           & 5600           &4.5       &--3.71    &1.0     &4.60\nl
\enddata 
\tablenotetext{a}{Microturbulent velocity (km s$^{-1}$)}
\end{deluxetable}
\end{center}

\clearpage
\begin{center}
\begin{deluxetable}{llr}
\tablecaption{ERROR BUDGET FOR CS~22876--032}
\tablehead{
\colhead{Error Source}                 &{$\Delta$}    & {$\Delta$[Fe/H]$_{\rm P}$} or\nl 
                                    &              & {$\Delta$(M$_{\rm S}$/M$_{\rm P}$)}\nl
        {(1)}                       &{(2)}         &{(3)}                     }
\startdata

{Impact on Systemic [Fe/H]:}\nl
Atmospheric parameters:\nl
~~T$_{\rm eff,P}$                   &130~K	        &0.10\nl
~~T$_{\rm eff,S}$                   &130~K	        &0.03\nl
~~Microturbulence, $\xi_{\rm P}$    &0.3~km~s$^{-1}$	&0.03\nl
~~Surface Gravity, log~$g_{\rm P}$  &0.3~dex		&0.01\nl\nl
Model Atmosphere Grid\nl
~~Temperature gradient              &\nodata    &0.10\nl\nl
Isochrones:\nl
~~Metallicity, log~Z	            &0.50~dex		&0.04\nl
~~Adopted Age		            &2~Gyr		&0.04\nl
~~Adopted Helium, Y		    &0.05	        &0.00\nl
\nl
{Impact on Mass Ratio (M$_{\rm S}$/M$_{\rm P}$):}\nl
Atmospheric parameters:            &as above		&0.02\nl
Isochrones:		           &as above		&0.03\nl

\enddata 

\end{deluxetable}
\end{center}

\clearpage
\begin{center}
\begin{deluxetable}{lrrr}
\tablecaption{SOURCES OF [Fe/H] DIFFERENCES}
\tablehead{
\colhead{Source}  &{NPB\tablenotemark{a}}  &{This work} &{[Fe/H]$_{\rm This}$--[Fe/H]$_{\rm NPB}$} \nl 
        {(1)}     &{(2)}                   &{(3)}       &{(4)}   
}

\startdata

T$_{\rm eff}$ (K)                       & 6000           & 6300                   & +0.24      \nl   
log~${g}$  (cgs)                         & 4.0            & 4.5                    & $\mid$$<$0.02$\mid$  \nl
$\xi$ (km s$^{-1}$)                     & 2.0            & 1.4                    & +0.05      \nl   
W$_{\lambda}$                           & As observed        &Corrected\tablenotemark{b}              & --0.03      \nl
log~${gf}$                               & Laboratory     & Laboratory             & $\mid$$<$~0.02$\mid$ \nl
Models                                  & Kurucz (1993) & Bell (1983)             & --0.12     \nl    
Model [Fe/H]                            & --4.0          & --2.0                  & +0.04      \nl   
log~(N$_{\rm Fe}$/N$_{\rm H}$)$_{\odot}$ & --4.33         & --4.50                 & +0.17      \nl

\enddata
\tablenotetext{a}{Norris et al.(1993)}
\tablenotetext{b}{Present measures are 0.08 dex smaller than those of NPB, but were multiplied by f$_{\rm 4300,P}$ = 1.28.}
\end{deluxetable}
\end{center}

\clearpage
\begin{center}
\begin{deluxetable}{llrrrrr}
\tablecaption{RELATIVE ABUNDANCES, [X/Fe], FOR CS~22876--032}
\tablehead{
\colhead{Element}  &{Feature}  &{log~(N/N$_{\rm H}$)$_{\odot}$} &{[X/Fe]$_{\rm P}$} &{s.e} &{[X/Fe]$_{\rm S}$} &{s.e}
\nl 
        {(1)}      &{(2)}      &{(3)}                         & {(4)}              &{(5)} &{(6)}              &{(7)}
}

\startdata
Mg  &3           & --4.42 &                               0.50 & 0.12&     0.51 &0.15\nl
Al\tablenotemark{a}  &3961 {\AA}        & --5.53 &      --0.45 & 0.12&   --0.42 &0.23\nl
Si  &3905 {\AA}  & --4.45 &                             --0.21 & 0.13&     0.29 &0.24\nl
Ca  &4226 {\AA}  & --5.64 &                              +0.01\tablenotemark{b} &0.13 &    --0.01\tablenotemark{b} &0.24\nl
Sc  &4246 {\AA}  & --8.90 &                          $<$  0.27 & 0.07&       ...&...\nl
Ti  &2           & --7.01 &                               0.08 & 0.11&     0.23 &0.23\nl
Cr  &4254 {\AA}  & --6.33 &                          $<$--0.22 & 0.04&      ... &...\nl
Mn  &4030 {\AA}  & --6.61 &                          $<$--0.18 & 0.04&$<$--0.26 &0.06\nl
Co  &3873 {\AA}  & --7.08 &                          $<$  1.25 & 0.04&      ... &...\nl
Ni  &3858 {\AA}  & --5.75 &                          $<$  0.00 & 0.04&      ... &...\nl
Sr  &4077 {\AA}  & --9.10 &                          $<$--0.65 & 0.05&      ... &...\nl
Ba  &4554 {\AA}  & --9.87 &                          $<$  0.38 & 0.07&      ... &...\nl
 
\enddata 

\tablenotetext{a}{From Al 3944 {\AA}, [Al/Fe]$_{\rm P}$ = --0.32 and
[Al/Fe]$_{\rm S}$ = --0.13, respectively.  As emphasized first by Arpigny \&
Magain (1983) this line suffers from potential blending with a feature of CH,
and is accordingly excluded from many analyses (e.g. Ryan et al. 1996),
including the present work.  For CS~22876--032 this may explain the higher
abundance derived from the $\lambda$~3944~{\AA} line in the secondary.}

\tablenotetext{b}{Following Magain (1988) and Ryan et al. (1991, 1996) we
have applied a correction of +0.18 dex to the abundances determined from Ca I
4226~{\AA}.  This line consistently yields a lower abundance than other
Ca I lines.}

\end{deluxetable}
\end{center}

\newpage
\begin{figure} [p]  
\centering \leavevmode 
\epsfxsize =\textwidth 
\epsfbox [100 100 500 550] {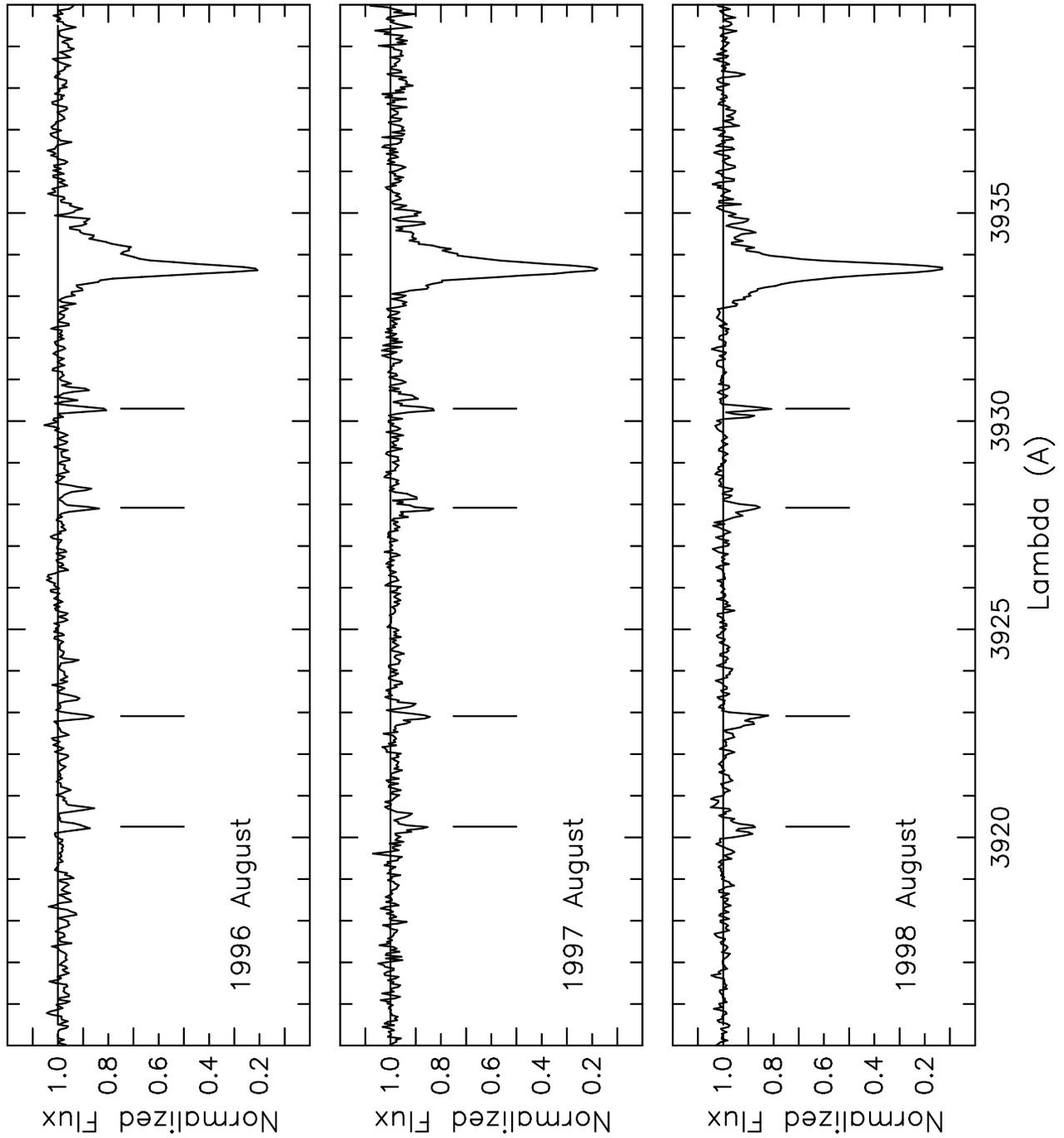}

\caption {Spectra of CS~22876--032 in the region of the Ca II K line.  The 
vertical lines indicate the positions of Fe I lines in the primary
component.}

\end{figure} 

\newpage
\begin{figure} [p]  
\centering \leavevmode 
\epsfxsize =\textwidth 
\epsfbox [100 100 500 550] {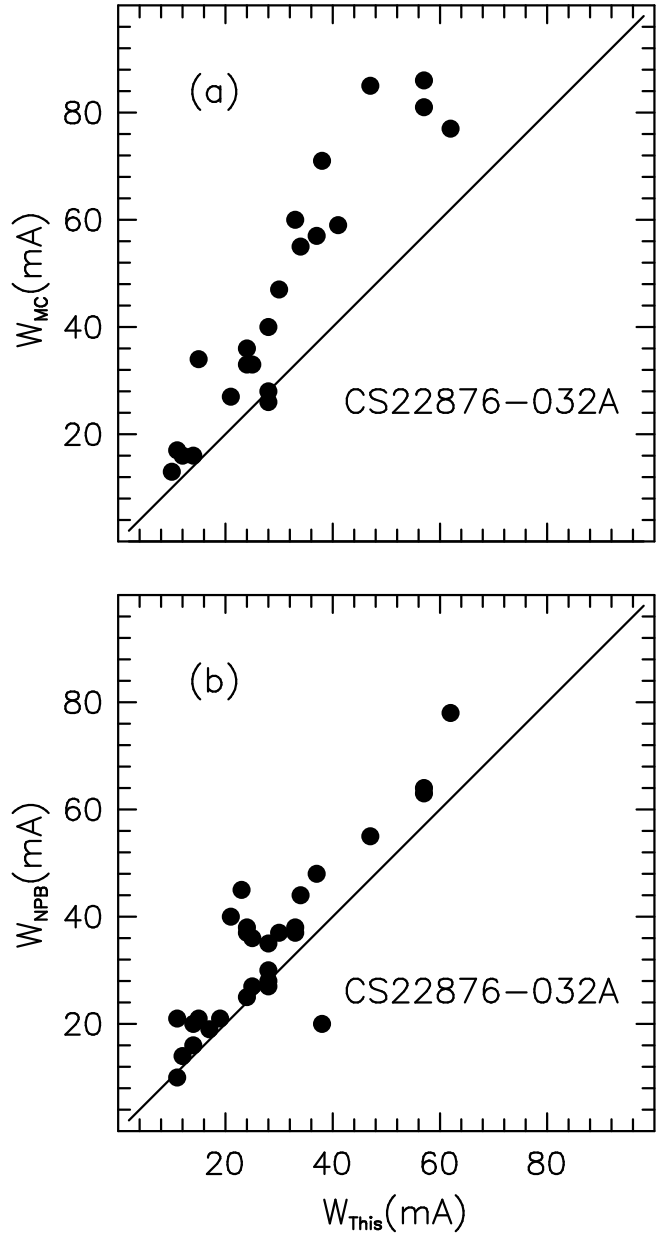}

\caption { Comparison of the equivalent widths presented here for
CS~22876--032A with those of (a) Molaro and Castelli (1990) and (b) Norris et
al. (1993).}

\end{figure} 

\begin{figure} [p]  
\centering \leavevmode 
\epsfxsize =\textwidth 
\epsfbox [130 100 490 540] {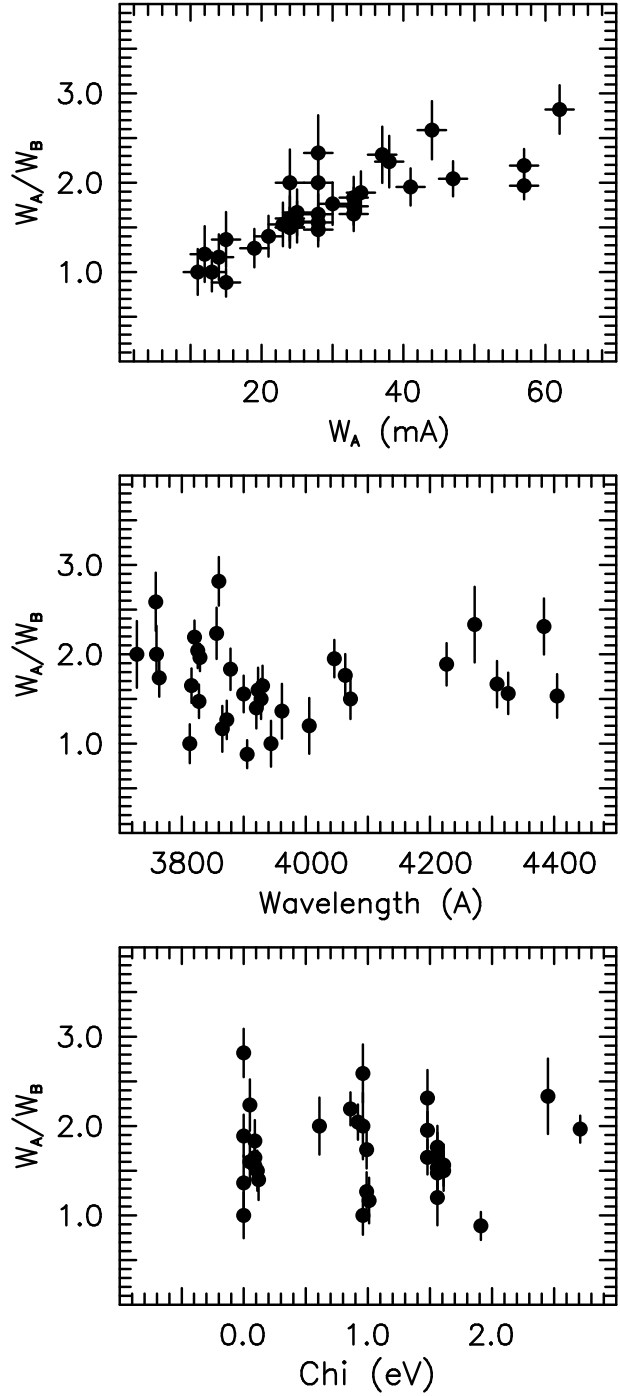} 

\caption {The dependence of W$_{\rm A}$/W$_{\rm B}$ on W$_{\rm A}$, wavelength, and
excitation potential.}

\end{figure} 

\newpage 
\begin{figure} [p] 
\centering \leavevmode 
\epsfxsize =\textwidth 
\epsfbox [130 100 490 540] {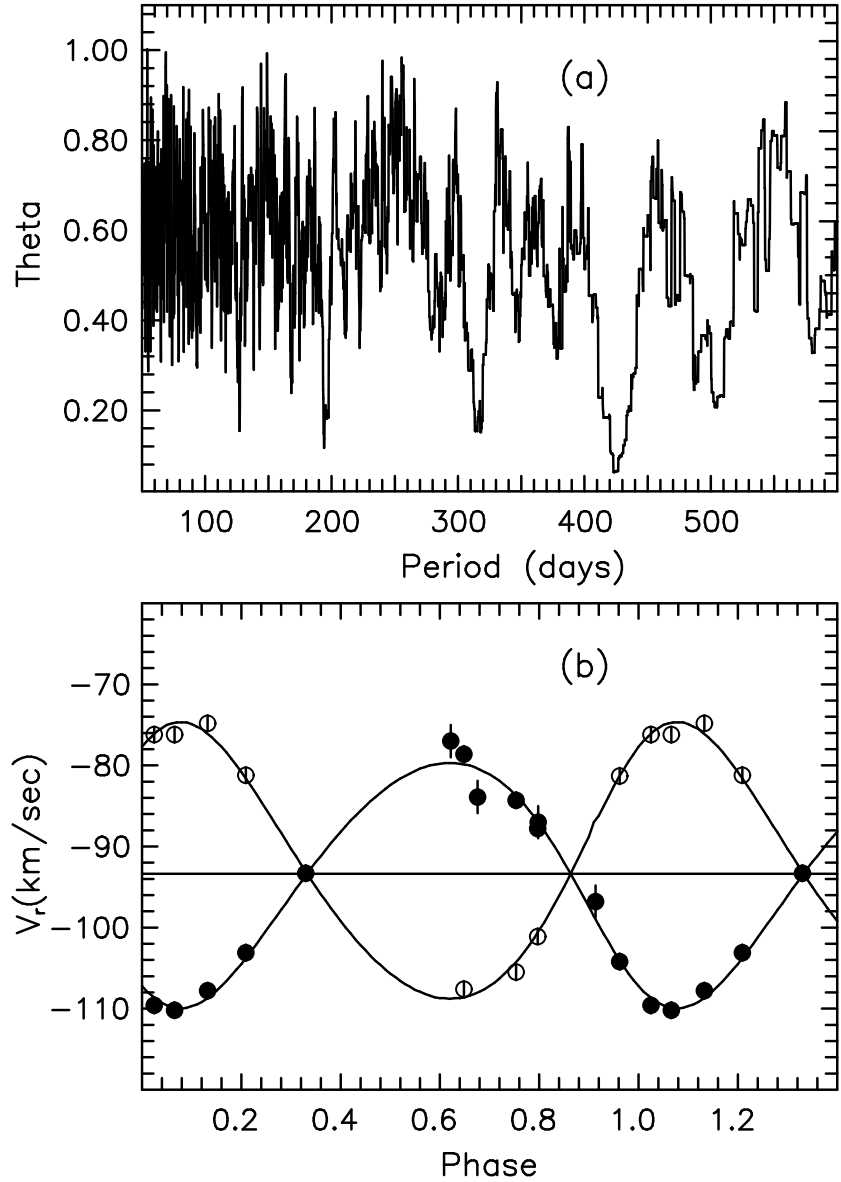} 

\caption {(a) Periodogram of radial velocities for CS~22876--032. ('Theta' is
as defined by Lafler \& Kinman (1965, Eqn. 1).) (b) Radial velocity curves
for CS~22876--032A (filled symbols) and CS~22876--032B (open symbols) for
period 424.71 days. 1$\sigma$ error bars, comparable in size to the symbols,
are also shown. The curves are determined by the elements in Table 3, while
the horizontal line represents the center-of-mass velocity of the system.}

\end{figure} 

\newpage 
\begin{figure} [p] 
\centering \leavevmode 
\epsfxsize =\textwidth 
\epsfbox [70 150 290 450] {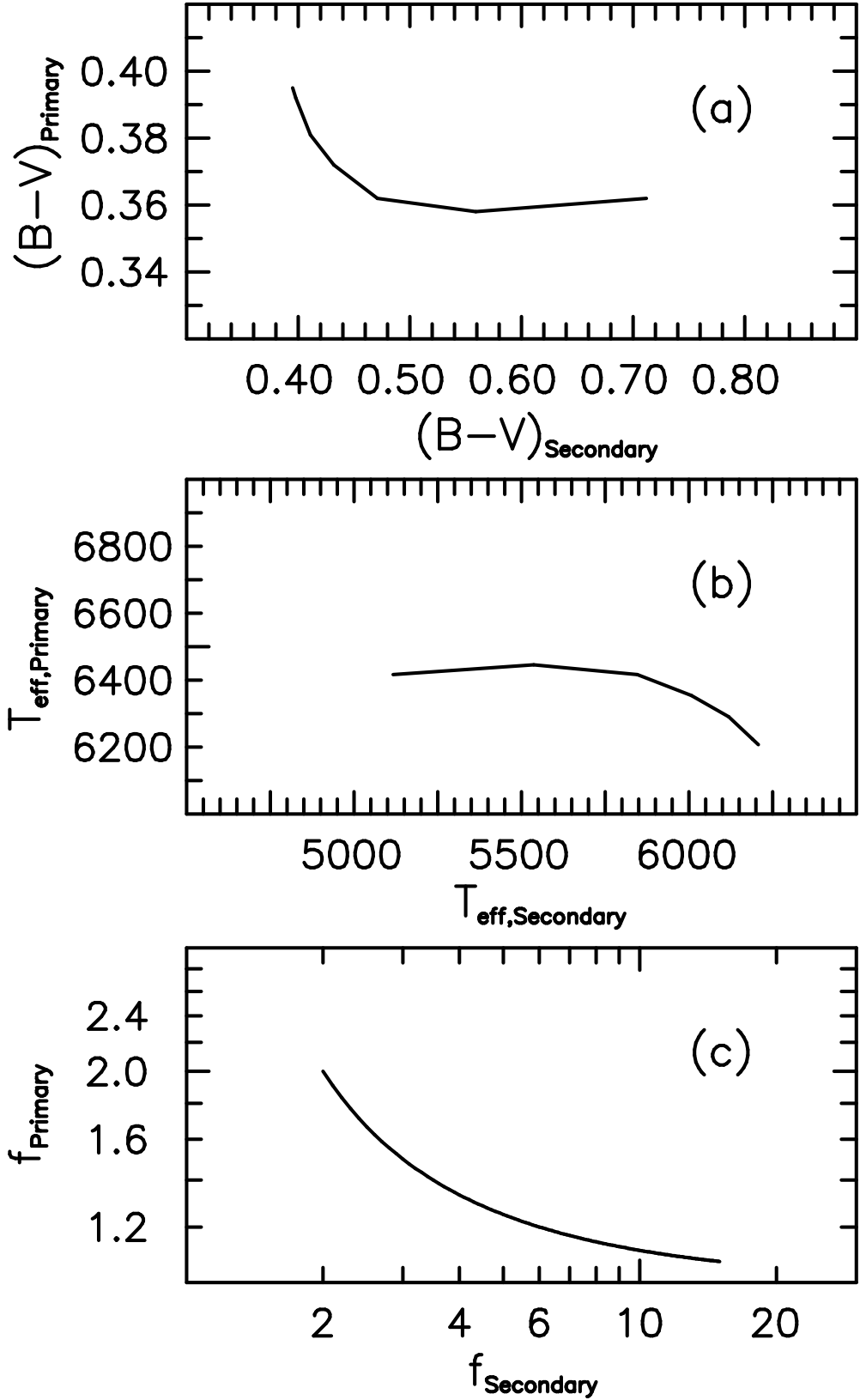} 

\caption {Locus of primary, secondary isochrone pairs of (a) {\it B--V}, (b)
T$_{\rm eff}$, and (c) equivalent width correction factors, f (applicable at
$\lambda$4300 {\AA}), consistent with {\it B--V} = 0.395 observed for
CS~22876--032A,B.}

\end{figure} 

\newpage 
\begin{figure} [p] 
\centering \leavevmode 
\epsfxsize =\textwidth 
\epsfbox [70 150 300 320] {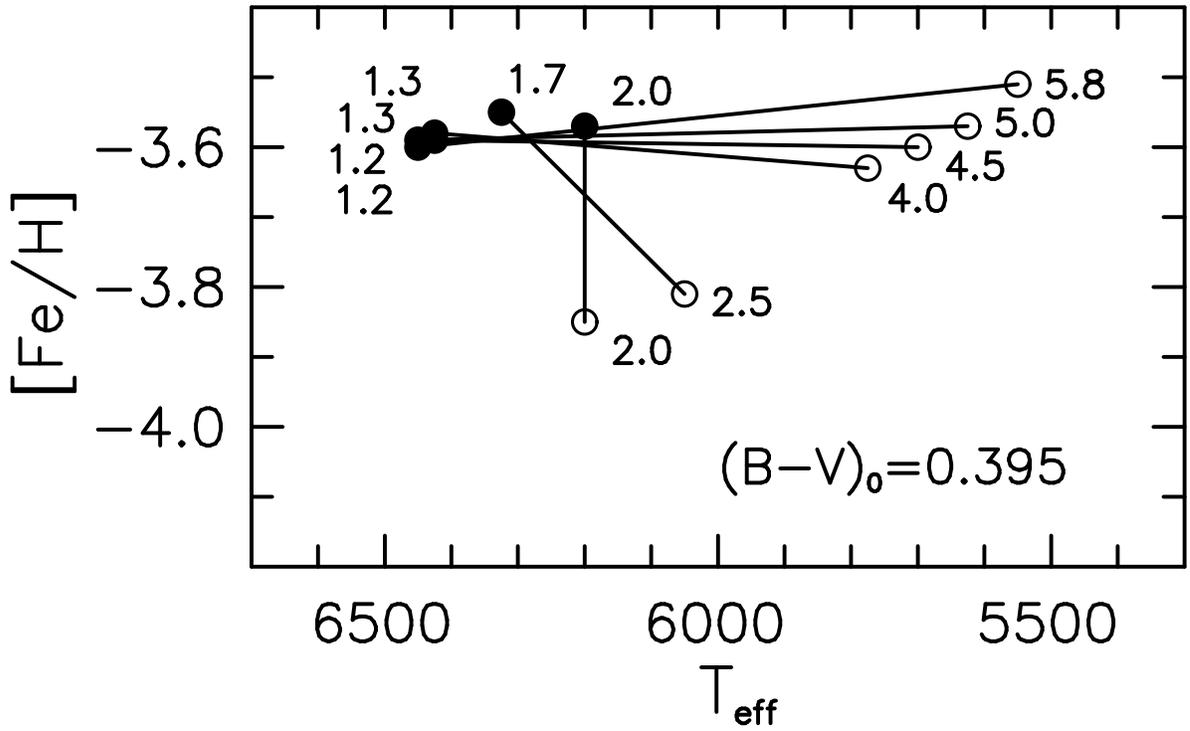} 

\caption {The dependence of [Fe/H] on T$_{\rm eff}$ for pairs of primary and
secondary isochrone components defined by the observed {\it B--V} and model
atmosphere analysis of the Fe I line strengths.  Filled and open symbols
refer to primary and secondary, respectively, and members of pairs are joined
by lines.  The equivalent width correction factors, f$_{\rm 4300}$, are
appended to the data points.  See text for discussion.}

\end{figure} 

\newpage 
\begin{figure} [p] 
\centering \leavevmode 
\epsfxsize =\textwidth 
\epsfbox [180 120 370 380] {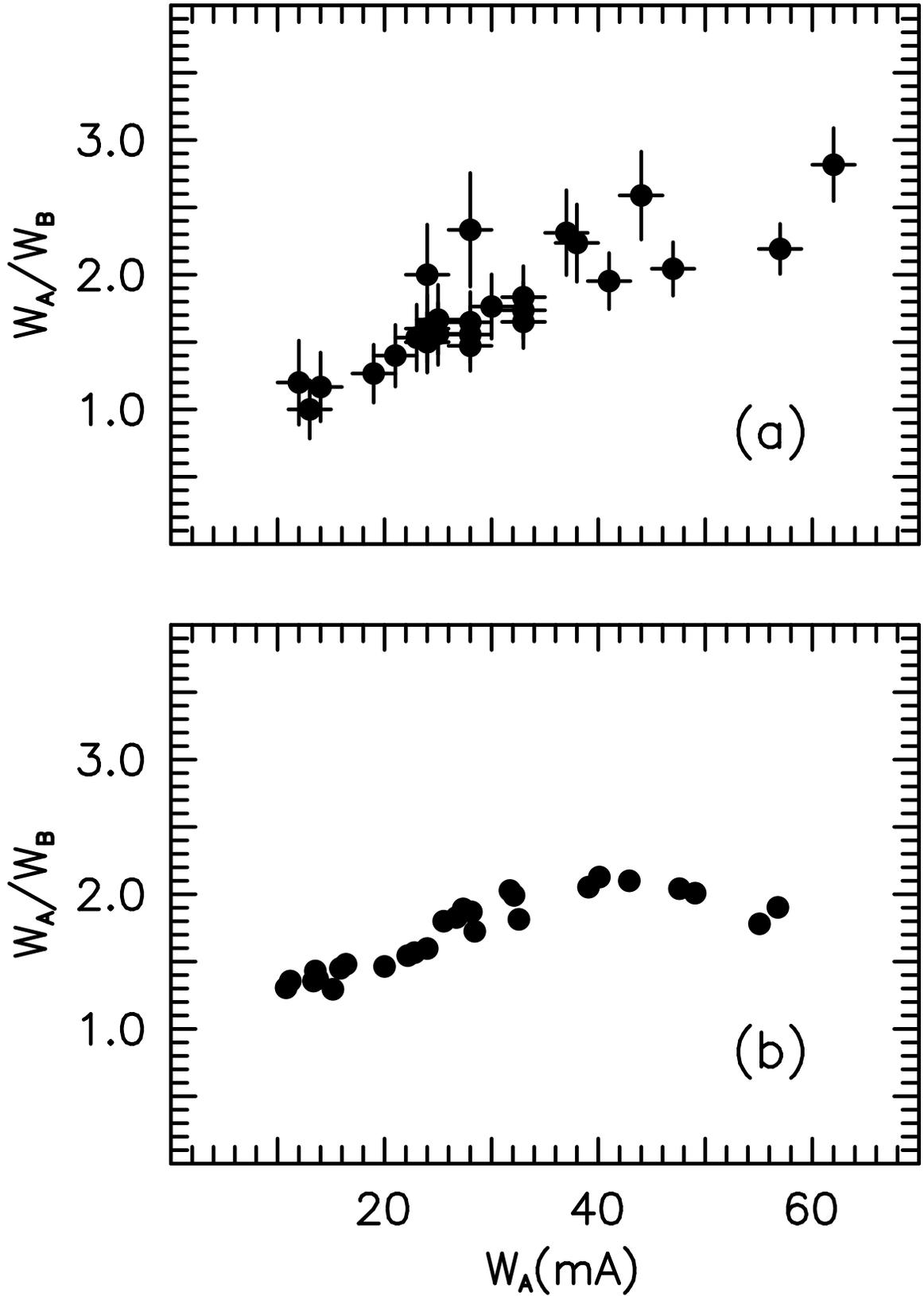} 

\caption {A comparison of (a) the observed and (b) predicted dependence of
W$_{\rm A}$/W$_{\rm B}$ on W$_{\rm A}$ for CS~22876--032.  In both panels
only Fe I lines stronger than 10 m{\AA} in both components in the composite
spectrum are included.}

\end{figure} 

\newpage 
\begin{figure} [p] 
\centering \leavevmode 
\epsfxsize =\textwidth 
\epsfbox [120 110 420 350] {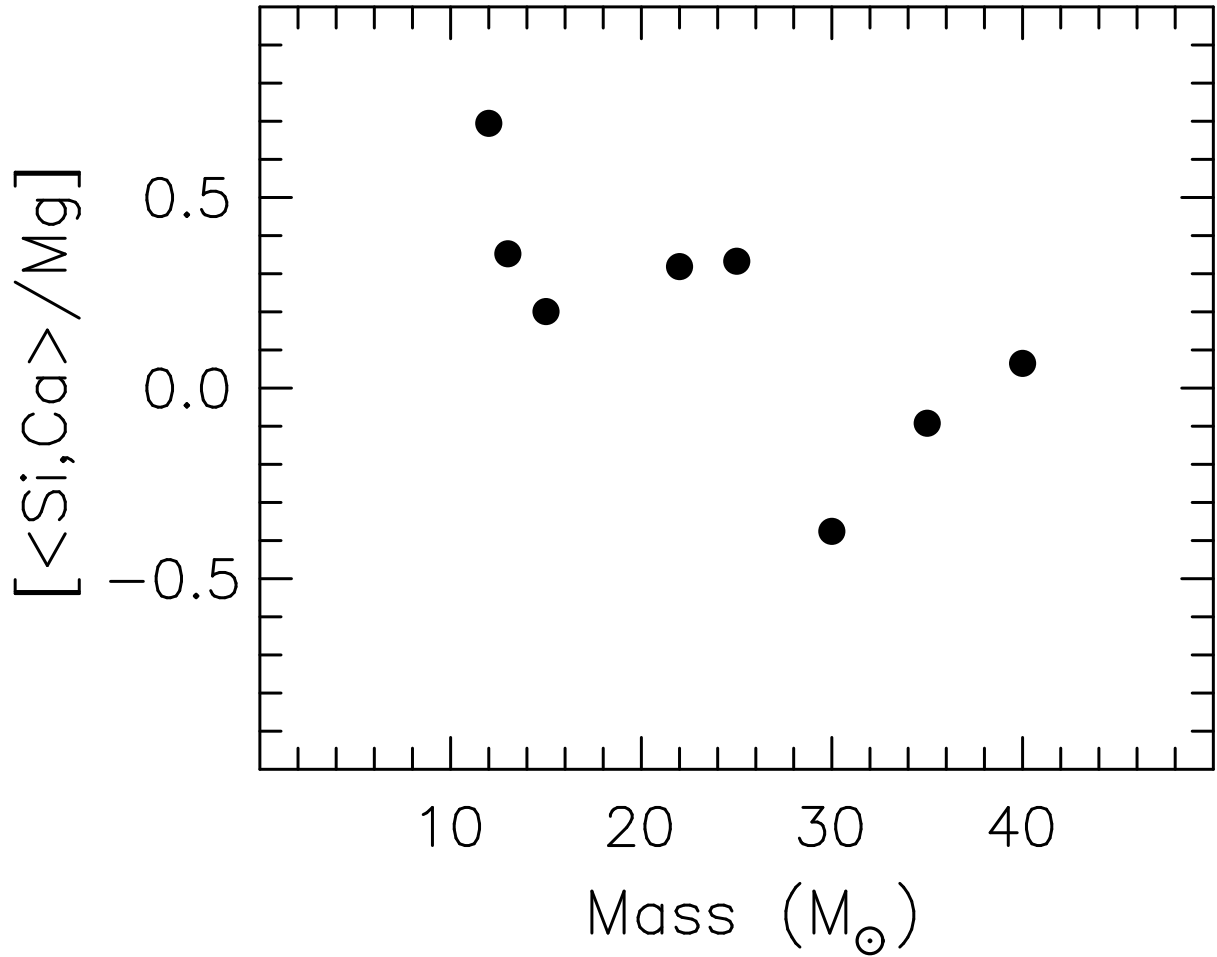} 

\caption {The dependence of [$\langle$Si,Ca$\rangle$/Mg] on supernova mass,
from the models of Woosley \& Weaver (1995).  See text for discussion.}

\end{figure} 

\newpage 
\begin{figure} [p] 
\centering \leavevmode 
\epsfxsize =\textwidth 
\epsfbox [120 110 410 350] {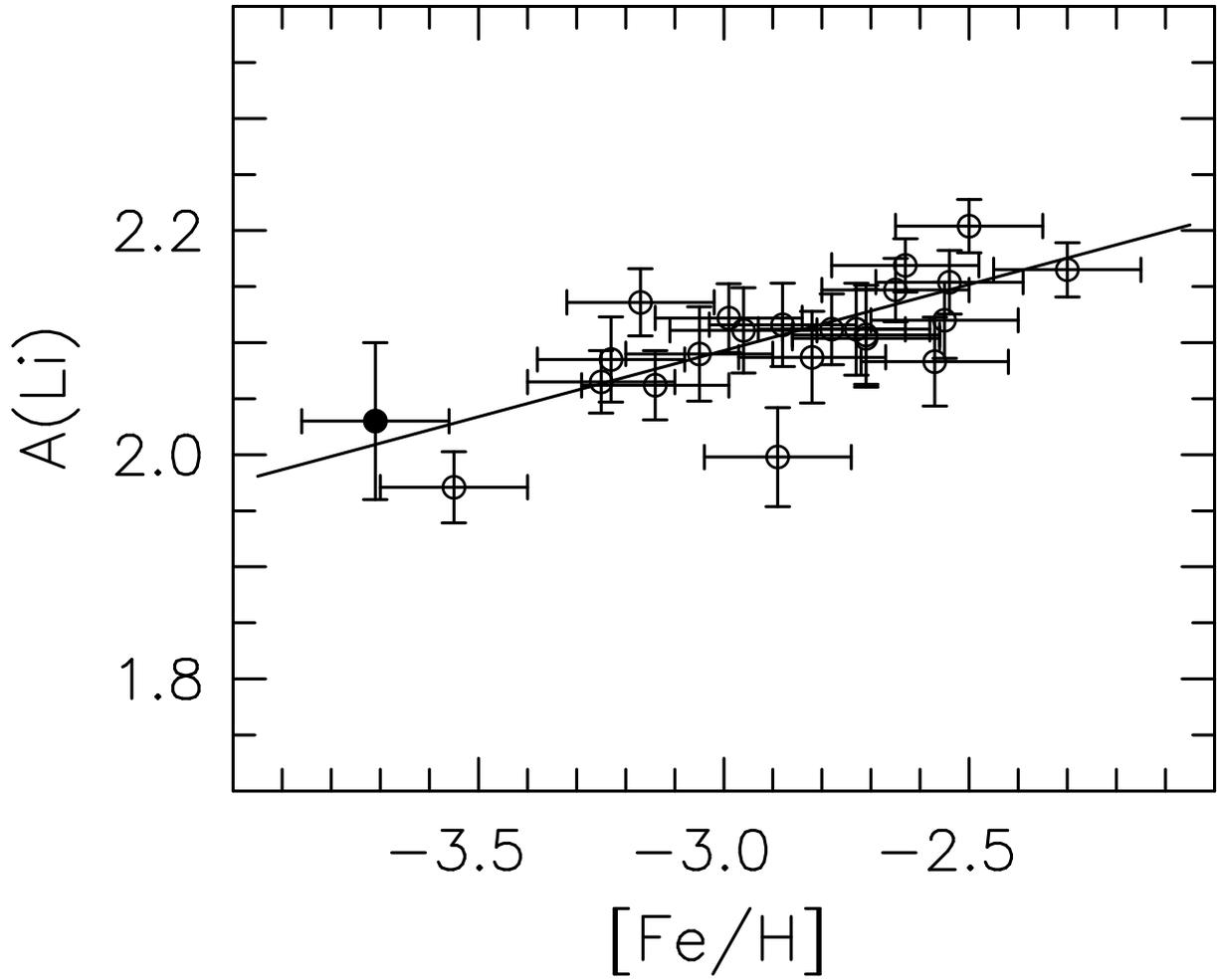} 

\caption {A comparison of the lithium abundance of CS~22876--032 (filled
symbol) with those of the unbiased near-main-sequence-turnoff sample of Ryan
et al. (1999) (open symbols).  The regression line A(Li) = 2.447 +
0.118$\times$[Fe/H] of Ryan et al. is also shown.}

\end{figure} 

\end{document}